\documentclass[twoside,11pt]{article}

\usepackage{xcolor}

\usepackage{subcaption}
\usepackage[frozencache=true,cachedir=minted-cache]{minted}
\usepackage[most]{tcolorbox}
\tcbuselibrary{minted, breakable} 

\newtcblisting[auto counter, number within=chapter, list inside=lol]{customminted}[3][]{
    listing engine=minted,
    listing only,
    breakable,
    minted language=python,
    minted style=default,
    minted options={breaklines,autogobble},
    colback=blue!5,
    colframe=black!75,
    enhanced,
    overlay={
        \begin{tcbclipinterior}
            \fill[black!15] (frame.south west) rectangle ([yshift=10pt]frame.north west);
        \end{tcbclipinterior}
    },
    title={Listing \thetcbcounter: #2},
    label={#3},
    #1
}
\usepackage{jmlr2e}

\ShortHeadings{AlgOS}{Salt and Gallagher}
\firstpageno{1}

\begin{document}

\title{AlgOS\: Algorithm Operating System}

\author{\name Llewyn Salt \email llewyn.salt@gmail.com \\
        \addr School of Electrical Engineering and Computer Science\\
        University of Queensland\\
        Brisbane, Australia
       \AND
       \name Marcus Gallagher \email marcusg@eecs.uq.edu.au\\
       \addr School of Electrical Engineering and Computer Science\\
       University of Queensland\\
       Brisbane, Australia}

\editor{}

\maketitle

\begin{abstract}
Algorithm Operating System (AlgOS) is an unopinionated, extensible, modular framework for algorithmic implementations. AlgOS offers numerous features: integration with Optuna for automated hyperparameter tuning; automated argument parsing for generic command-line interfaces; automated registration of new classes; and a centralised database for logging experiments and studies. These features are designed to reduce the overhead of implementing new algorithms and to standardise the comparison of algorithms. The standardisation of algorithmic implementations is crucial for reproducibility and reliability in research. AlgOS combines Abstract Syntax Trees with a novel implementation of the Observer pattern to control the logical flow of algorithmic segments. 
\end{abstract}


\section{Introduction}
In algorithmic research, particularly where stochasticity is involved, there have been concerns for reproducibility and repeatability~\citep{islam2017reproducibility, henderson2018deep, alahmari2020challenges,lynnerup2020survey,lopez2021reproducibility,albertoni2023reproducibility}. Reinforcement learning algorithms, for example, are notoriously difficult to reproduce due to the inherent randomness in the environment and the algorithm itself. The same algorithm can produce different results on different runs, even with the same hyperparameters to the point that we see tuning of random seeds~\citep{henderson2018deep}. Without mathematical formalisation, we rely on empirical results to evaluate the performance of an algorithm~\citep{lopez2021reproducibility}. If these results are not repeatable due to stochasticity, a lack of completeness in the methodology, versioning issues with software, or not fully describing the hyperparameters used then it becomes difficult to reproduce and verify the results~\citep{lynnerup2020survey}. Alahmari et al. discuss that even fixing the random seed in some libraries will not guarantee reproducibility~\citep{alahmari2020challenges}. Henderson et al. however suggest that fixing the random seed unless part of the algorithm doesn't provide results that give insight into the performance of the algorithm and that averaging across multiple runs to get a more accurate estimate of the distribution of results is a better approach~\citep{henderson2018deep}. There have been many modular reinforcement learning libraries developed to attempt to standardise implementations to assist in benchmarking popular algorithms~\citep{baselines,stable-baselines,stable-baselines3,liang2018rllib,castro18dopamine,hoffman2020acme,tianshou}. However, these libraries are often opinionated and do not provide methods for easily integrating new algorithms, nor do they provide a standardised way to store and compare results. 

We develop Algorithm Operating System AlgOS to address these issues. AlgOS is a self-organising modular framework written in Python that is developed with the sole assumption that the algorithms to be developed can be represented as a cyclic or acyclic graph. AlgOS offers multi-tiered modularity, dependency injection, hyperparameter optimisation integration, and a centralised database for logging experiments and studies. The aggregation of these features provides a framework that encourages experimental standardisation and reproducibility. A researcher can reuse aspects of another's work without needing to reproduce another's code with the potential for introducing their own logical errors. Additionally, due to the generic nature of the framework and the associated database, researchers could include their data as well as their code when presenting results. As the framework is standardised, this would enable verification of results by other researchers and a more iterative approach to research, as researchers would no longer have to invest time in reimplementing algorithms. We develop AlgOS with a core OS that is extensible to any algorithm that can be represented as a graph. We demonstrate the utility of AlgOS by implementing a reinforcement learning module that wraps existing implementations and allows the injection of additional algorithmic logic. In summary, the core contributions of AlgOS are:
\begin{enumerate}
    \item A novel implementation self-organising algorithmic logical flow that utilises:
    \begin{enumerate}
        \item Abstract Syntax Trees to parse code structures input/output features~\citep{welty1997augmenting,lin2021improving}.
        \item A novel implementation of the observer pattern~\citep{gamma1995design} with threadsafe gated state access.
        \item A factory pattern that utilises Subcomponent for dependency injection~\citep{sun2022measuring}.
        \item Hyperparameter definition at the Subcomponent and Component level which:
            \begin{enumerate}
                \item Automates the optimisation process.
                \item Allows for easy hyperparameter definition.
                \item Makes hyperparameter optimisation consistent across studies.
            \end{enumerate}
    \end{enumerate}
    \item Database logging to standardise and centralise experimental data collection.
    \item Automated argument parsing for generic Command Line Interface (CLI) scripts.
    \item Automated class registration to enable easy extension of the framework.
    \item Heavy integration with Optuna~\citep{optuna_2019} for hyperparameter optimisation.
    \item Automated hyperparameter optimisation CLI, remote and local.
    \item A plotting module for database data extraction and visualisation.
\end{enumerate}

\section{Design and Implementation}
\begin{figure}[htbp]
  \begin{minipage}[t]{0.6\textwidth}
      \centering
      \includegraphics[height=6.5cm]{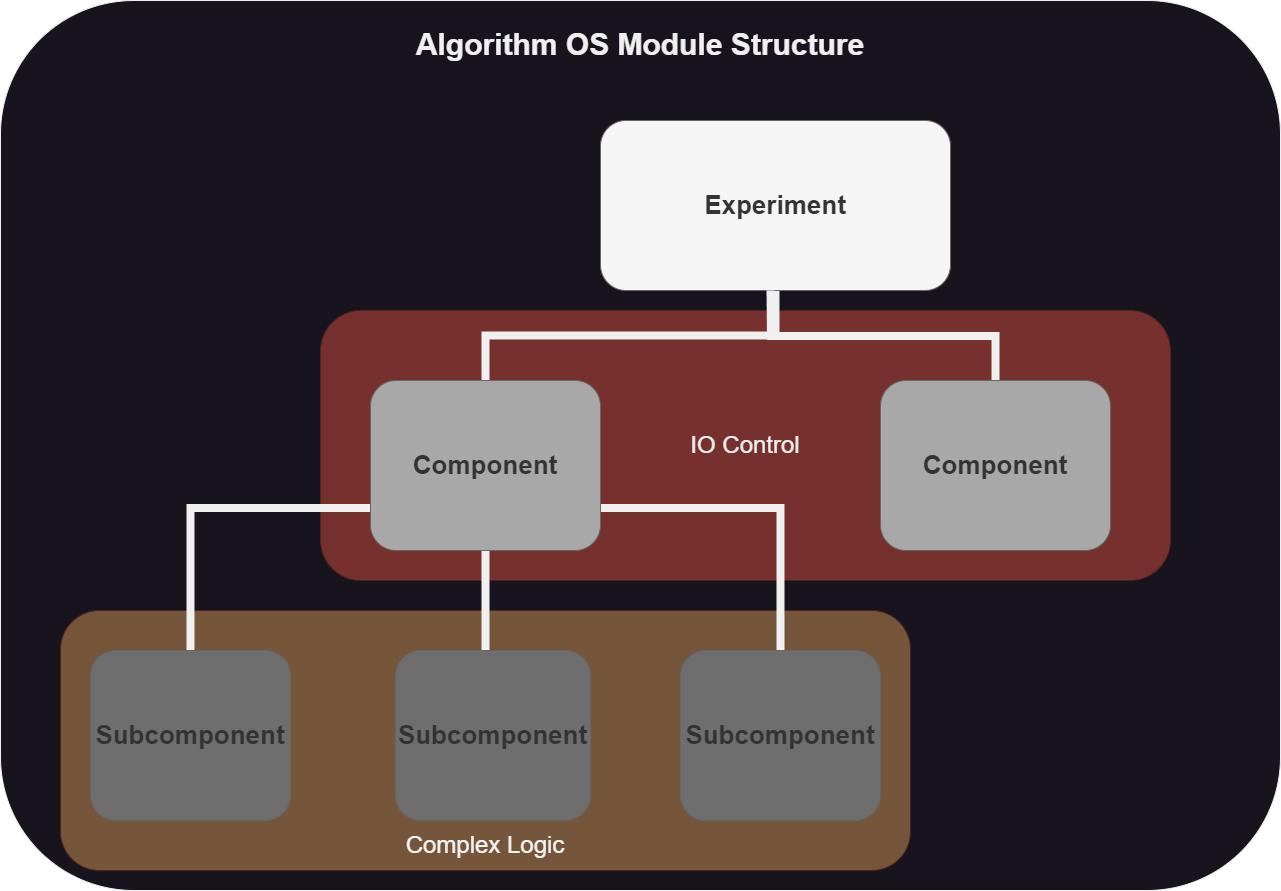}
      \caption{Modular Organisation of AlgOS.}
      \label{fig:algosorganisation}
  \end{minipage}
  \hfill
  \begin{minipage}[t]{0.35\textwidth}
      \centering
      \includegraphics[height=6.5cm]{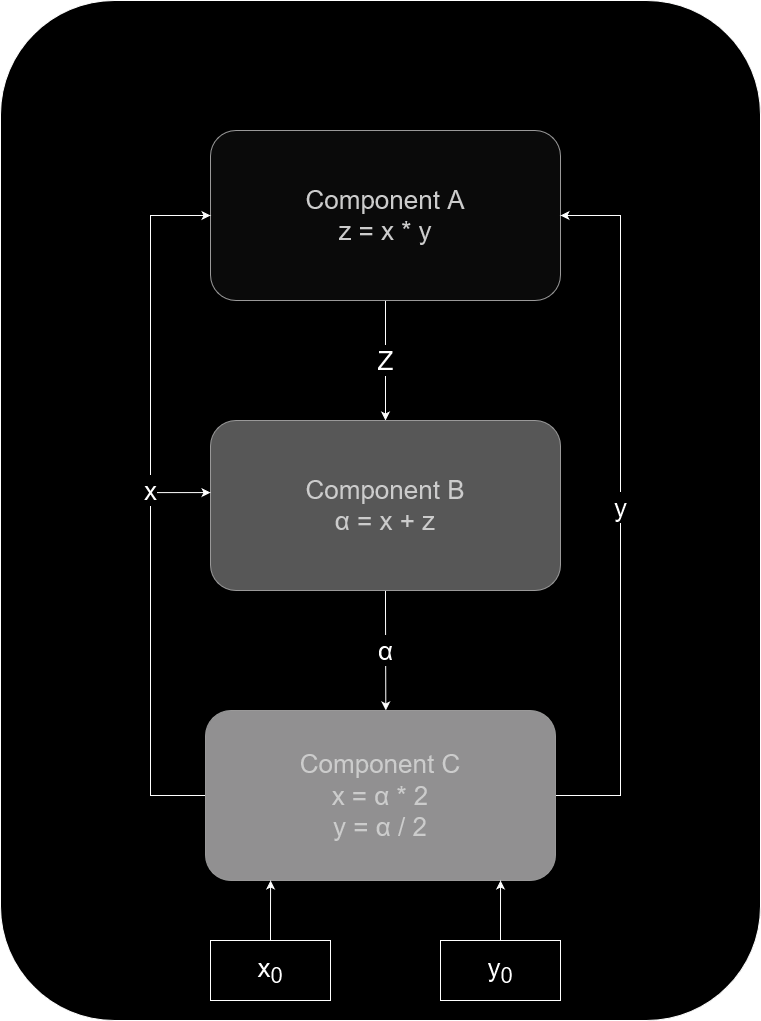}
      \caption{Toy Example.}
      \label{algos:arch:fig:1}
  \end{minipage}
\end{figure}

AlgOS is designed to have three tiers of modularity: Experiment, Component, and Subcomponent. Each tier can have a one-to-many relationship with the tier below, however an experiment needs only one component and a component can have no subcomponents. Figure~\ref{fig:algosorganisation} shows the relationships of these modules. Both components and subcomponents have hyperparameters defined that enables both automated hyperparameter optimisation and automated argument parsing for a standardised experimental command line interface.

As shown in Figure~\ref{fig:algosorganisation}, the components facilitate the logical flow of the algorithm through I/O control. This is done by using an Abstract Syntax Tree to parse the logic of the code written within the component which utilises an IO map defined within a concrete component that determines the namespace of the input and output variables. The inputs and outputs are modelled using a heavily modified Observer pattern. That is the inputs are Observers, and the outputs are Subjects. The Observer and Subjects are defined as pseudo singletons based on their namespace and owner and namespace respectively. If a subject is created with the same namespace an error is raised, if an observer is created with the same namespace and owner then the same Observer is returned with some internal logic. The combination of the modified Observer pattern and the AST stringently controls the logical flow of the algorithm by ensuring that subjects can only set their values when the observers have observed the current value and observers can only access their value when the subjects have set the value. This is enabled having the components class inherits from threading to enable asynchronous logical progression. If the graph is incomplete or there is a logical error that causes a deadlock then the code will time out. 

Figure~\ref{algos:arch:fig:1} we can see three components with simple functionality. Component A requires inputs x and y and outputs z, component B takes inputs x and z and outputs $\alpha$, and component C takes input $\alpha$ and outputs x and y. In this instance component C will initiate the logical flow of the algorithm by setting the initial states of x ($x_0$) and y($y_0$). This will trigger component A to set z which will in turn trigger component B to set $\alpha$. Once $\alpha$ is set component C will set x and y. This will repeat until some terminating condition for the algorithm. Using the I/O map we could map alpha to a different namespace and then intercept its value in component D before transmitting to components A and B, or transmit the original value to one and the changed value to another. The power of this architecture is that components are agnostic of each other and work independently of each other. 

\textbf{Hyperparameter Optimisaton:} Optuna is embedded in the framework through the automated hyperparameter collection by exploiting class methods and metaclassing. This enables the user to specify bounds for each subcomponent and have this consistently applied across experiments and studies. The CLI can run the optimisation remotely using SLURM~\citep{yoo2003slurm, simakov2018slurm} or locally and supports multiple parallel processes. Optuna offers a web interface for visualising the optimisation process and results. Additionally, as Optuna uses SQL databases to store the results we fully integrate with the Optuna database and provide a plotting module to extract and visualise the data.

\textbf{Database Logging:} We provide a complete database logger utilising SQLAlchemy, currently supporting PostgreSQL and SQLite. The logger is run as a separate process so that it doesn't block the algorithmic execution, it supports a local SQLite database as a backup if connectivity to the remote database is lost ensuring data integrity. It commits in bulk chunks to reduce overhead and can be vertically scaled to handle large amounts of data by expanding the number of database workers running. A simple interface is provided to all components so that the user can log data using logger.record(tagname, value) which mimics the Tensorboard interface~\citep{tensorboard2025}. This allows the user to wrap libraries that use Tensorboard and replace the logger with the AlgOS logger.  

\textbf{Visualisation:} As the database is centralised, we provide a plotting module that can extract data from the database and visualise it using Matplotlib~\citep{Hunter:2007} and Seaborn~\citep{Waskom2021}. By automating result generation and guiding standardised data extraction, this approach enhances methodological consistency and supports reproducible research practices.

\textbf{Dependency Injection:} We provide a \mintinline{python}|RegistryMeta| metaclass that registers all classes that form part of each modular level with their own registry. This allows for easy extension of the framework by simply defining a new class and inheriting from the appropriate superclass. The registry is used to formalise the hyperparameters and input arguments required to build said classes, which then interfaces with a factory pattern to allow for dependency injection when building components or experiments.

\section{Conclusion}

We present a feature rich generalised framework, AlgOS, for algorithmic research that standardises implementation, optimisation, and logging. Through some novel algorithmic and architectural choices, AlgOS provides a modular, extensible, and flexible framework to enable researchers to iteratively extend algorithms. This should reduce the overhead of implementing new algorithms and standardise the way they are compared. Hopefully, this ameliorates the reproducibility and repeatability issues that have engendered concern in the algorithmic research community.

\newpage
\appendix
\section*{Appendix A. Simple Code Example}
\label{app:simplecode}
We can create a components A, B, and C from Figure~\ref{algos:arch:fig:1} as follows:
\begin{customminted}{Component Definition}{code:compdef}
  class ComponentA(RunnerDummyInterface):
      _io_map = {"x":"x", "y":"y", "z":"z"} 
      def initiate_sequence(self):
          pass
      def step(self):
          temp = self.x * self.y
          self.z = temp
  
  class ComponentB(RunnerDummyInterface):
      _io_map = {"x":"x", "z":"z", "alpha":"alpha"}
      def initiate_sequence(self):
          pass
      def step(self):
          self.alpha = self.x + self.z
  
  class ComponentC(RunnerDummyInterface):
      _io_map = {"x":"x", "y":"y", "alpha":"alpha"} 
      def initiate_sequence(self):
          self._x.initialise_state(1)
          self._y.initialise_state(1)
      def step(self):
          temp = self.alpha*2
          self.x = temp
          temp = self.alpha/2
          self.y = temp
\end{customminted}

The io map is used to define the namespace of the input and output variables. The keys are the internal namespace and the values are the external namespace, which means that we can intercept the values of a variable before it is transmitted to another component. 

A container class \mintinline{python}|ComponentCollection| is provided to tie all the components together and implement the thread join/run boilerplate.
\begin{customminted}{Example Naive Experiment with OS module}{code:exampleAlgOS}
A = ComponentA()
B = ComponentB()
C = ComponentC()
component_collection = ComponentCollection(A = A, B = B, C = C)
component_collection.run()
\end{customminted}
The above code will execute the components as specified by the \mintinline{python}{RunnerDummyInterface} superclass. As the user defines the run function of the components this provides a modular framework. 

If we then desired to intercept a value and transform it before it is transmitted to another component we could implement ComponentD as follows:
\begin{customminted}{ComponentD Definition}{code:compDdef}
  class ComponentD(RunnerDummyInterface):
      _io_map = {"alpha":"alpha", "beta":"beta"}
      def initiate_sequence(self):
          pass
      def step(self):
          self.beta = self.alpha*2
\end{customminted}

We would then need to update the io map of component C to adjust the external namespace of alpha to beta, we can do this in the experimental setup:
\begin{customminted}{Example Naive Experiment with Intercept with OS module}{code:example2AlgOS}
A = ComponentA()
B = ComponentB()
C = ComponentC(io_map = {"x":"x", "y":"y", "alpha":"beta"})
D = ComponentD()
component_collection = ComponentCollection(A = A, B = B, C = C, D = D)
component_collection.run()
\end{customminted}

We show the modularity, extensibility, and flexibility of AlgOS as the code for ComponentC remains unchanged, yet the logical flow of the algorithm has been altered. 

\begin{figure}[htbp]
  \begin{center}
    \includegraphics[width=0.5\textwidth]{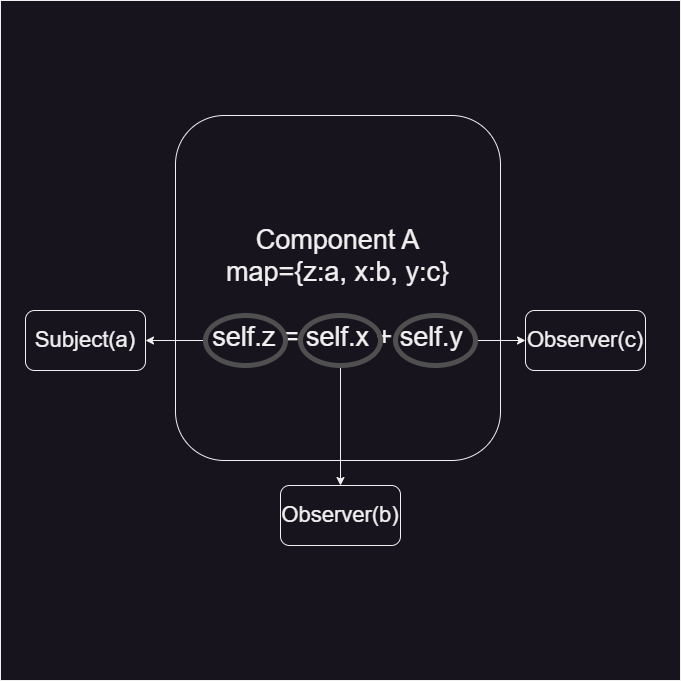}
    \caption{Component I/O.}
    \label{fig:componentio}
  \end{center}
\end{figure}

Figure~\ref{fig:componentio} shows how the subjects (outputs) and observers (inputs) are captured within a component visually. If the variables are defined within the keys of the io\_map then any class attributes are assumed to be part of the I\/O space. Any code that is assigned values becomes a subject and any variables that access a variable becomes an observer.

\section*{Appendix B. Factory Examples}
\label{app:factory}
As stated in the body of the paper we use a factory pattern to enable dependency injection. Assume that we have defined some components and subcomponents as follows:
\begin{customminted}{Component Factory}{code:factory}
  class ToyComponentFactory(AbstractComponentFactory):
    _register = []
    def build(self):
      if not hasattr(self, "_extra_args"):
        self._extra_args = {}
      input_args = {**self._extra_args, **get_class_args(register[0], self._experiment._exp_args)}
      self._components[register[0].__name__[-1]] = self.register[0](**input_args)
  class ComponentAFactory(ToyComponentFactory):
    _register = [ComponentA]
  class ComponentBFactory(ToyComponentFactory):
    _register = [ComponentB]
  class ComponentCFactory(ToyComponentFactory):
    _register = [ComponentC]
    def build{self}:
      self._extra_args = {"io_map": {"x":"x", "y":"y", "alpha":"beta"}}
      super().build()
  class ComponentDFactory(ToyComponentFactory):
    _register = [ComponentD]
  class ComponentFFactory(ToyComponentFactory):
    _register = [ComponentF, SubcomponentA, SubcomponentB]
    def build{self}:
      for T in self._register[1:]:
        if isinstance(T, SubcomponentA):
          sub_compA = T{**get_class_args(T, self._experiment._exp_args)}
        elif isinstance(T, SubcomponentB):
          sub_compB = T{**get_class_args(T, self._experiment._exp_args)}
        self._exp_args = {'sub_compA': sub_compA, 'sub_compB': sub_compB}
      super().build()
\end{customminted}

This provides the interface for the user to define the components and subcomponents to be used in the experiment. Assume \mintinline{python}|ComponentF| utilises subcomponents, \mintinline{python}|SubcomponentA| and \mintinline{python}|SubcomponentB|, then the user can define the factory as above. The \mintinline{python}|get_class_args| function is a helper function that extracts the keyword arguments for a given component or subcomponent and returns them as a dictionary. This enables us to exchange the classes in the registry for \mintinline{python}|ComponentFFactory| and algorithmic segments executed by the subcomponents. This demonstrates the ease with which a researcher can iteratively extend an algorithm with limited code changes.

Defining an experiment is similarly simple it leverages instead the \mintinline{python}|AbstractExperiment| class and the factory interface to build the components and subcomponents:

\begin{customminted}{Example Experiment Definition}{code:expdef}
  class ToyExperiment(AbstractExperiment):
    _register = [ComponentAFactory, ComponentBFactory, ComponentCFactory, ComponentDFactory]
  class ToyExperimentF(AbstractExperiment):
    _register = [ComponentAFactory, ComponentBFactory, ComponentCFactory, ComponentFFactory]
\end{customminted}

The experiment factory coupled with the component factory allows for easy extension of existing algorithms and significant reuse of logic. Components are agnostic of each other and progressed based on input/output gating, which means that the user does not need to understand the internal workings of the components to extend the algorithm. 

\section*{Appendix C. Hyperparameter Optimisation}
\label{app:hyperparam}
The hyperparameter optimisation is automated through the use of Optuna. The user can define the hyperparameters at the subcomponent level and the component level. The user can define the hyperparameters in the following way:
\begin{customminted}{Hyperparameter Definition}{code:hyperparam}
  class ComponentF(RunnerDummyInterface):
    _io_map = {"alpha":"alpha", "beta":"beta"}
    def __init__(self, sub_classA, sub_classB, *args, **kwargs):
        self._sub_classA = sub_classA
        self._sub_classB = sub_classB
        super().__init__(*args, **kwargs) 
    def initiate_sequence(self):
        pass
    def step(self):
        alpha = self.alpha
        intermediate = self._sub_classA(alpha)
        self.beta = self._sub_classB(intermediate)
        
    @classmethod
    def set_up_hyperparameters(cls):
        pass

  class SubcomponentA(AbstractParametered):
    def __init__(self, scaler:float = 0.1, *args, **kwargs):
        super().__init__(*args, **kwargs)
        self.scaler = scaler
    
    def __call__(self, alpha):
        return alpha * self.scaler

    @classmethod
    def set_up_hyperparameters(cls):
        cls._hyperparameters["scaler"].bounds = (0.1, 1.0)

  class SubcomponentB(AbstractParametered):
    def __init__(self, scaler:float = 0.1, *args, **kwargs):
        super().__init__(*args, **kwargs)
        self.scaler = scaler
    
    def __call__(self, intermediate):
        return intermediate * self.scaler

    @classmethod
    def set_up_hyperparameters(cls):
        cls._hyperparameters["scaler"].bounds = (0.2, 0.5)
\end{customminted}

\mintinline{python}|SubcomponentA| and \mintinline{python}|SubcomponentB| inherit from the subcomponent class \mintinline{python}|AbstractParametered| are defined with a hyperparameter \mintinline{python}|scaler| that is bounded between 0.1 and 1.0 and 0.2 and 0.5 respectively. The subcomponent class provides the abstract method \mintinline{python}|set_up_hyperparameters| which are collected by the components and finally the experiments to be passed to the optimisation commandline interface script. If a keyword argument is not provided bounds in the \mintinline{python}|set_up_hyperparamters| method then they are not optimised and instead the default value is used or they can be specified in the command line interface invocation. 

\section*{Appendix D. Database Logging}
\label{app:database}
The database logging is implemented using SQLAlchemy and is run as a separate process to the algorithmic execution, every component is provided with a ProxyLogger instance that associates all record invocations with the component. We can setup logging within components as follows:
\begin{customminted}{Database Logging}{code:dblog}
  class ComponentA(RunnerDummyInterface):
    _io_map = {"x":"x", "y":"y", "z":"z"} 
    def initiate_sequence(self):
        pass
    def step(self):
        temp = self.x * self.y
        self.z = temp
        self.logger.record("z", temp)
\end{customminted}

\begin{figure}[htbp]
  \begin{center}
    \includegraphics[width=0.8\textwidth]{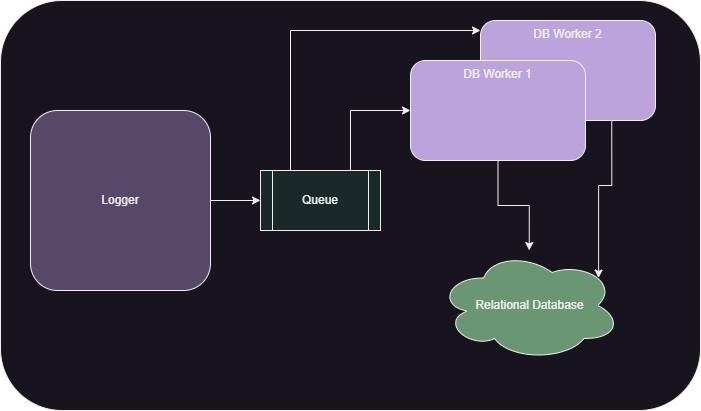}
    \caption{Database Logging Process.}
    \label{fig:dblog}
  \end{center}
\end{figure}

If we want to log in subcomponents, we can pass the logger to the subcomponent as an input argument. The database logger is a separate process to avoid slowing down the algorithmic execution, see Figure~\ref{fig:dblog} for a visual representation of the database logging process.

\section*{Appendix E. Visualisation}
\begin{figure}
  \begin{center}
    \includegraphics[width=0.6\textwidth]{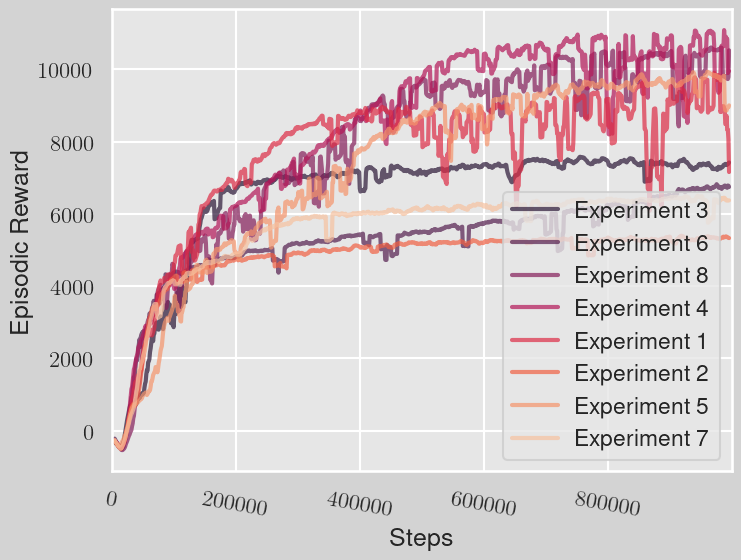}
    \caption{The reward in some reinforcement learning experiments.}
    \label{fig:plot1}
  \end{center}
\end{figure}

\begin{figure}
  \begin{center}
    \includegraphics[width=0.6\textwidth]{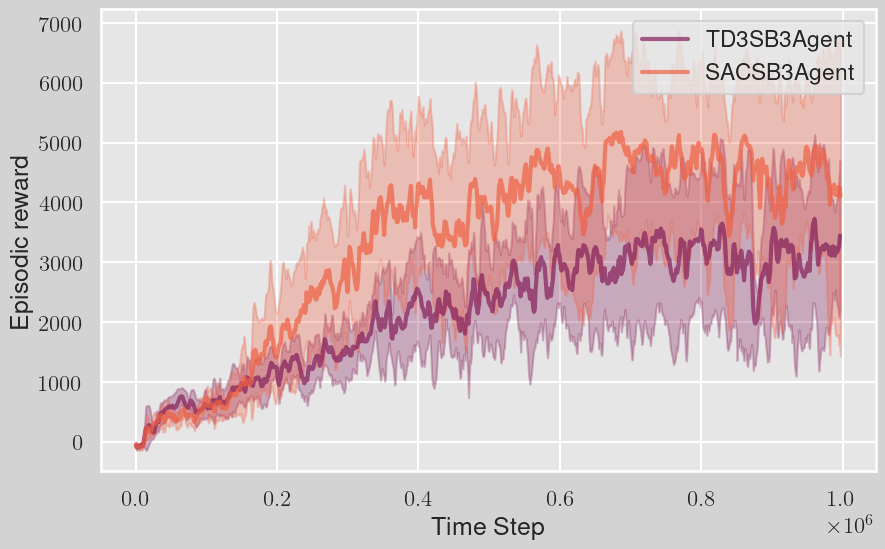}
    \caption{A comparison of the performance of two reinforcement learning algorithms.}
    \label{fig:plot2}
  \end{center}
\end{figure}

The visualisation module offers interfaces for extracting experimentation data from the database and to make generic plots. It also utilises seaborn and matplotlib to enable a variety of visualisations and a way to standardise teh look and feel of plots. The methods are designed with the same modular philosophy of the rest of the framework. Some example plots from our module are shown in Figure~\ref{fig:plot1} and Figure~\ref{fig:plot2}.

See the plotter.py file in the repository for more plotting examples. The visualisation module is designed to be easily extended and to provide a standardised way to extract data from the database and visualise it.

\vskip 0.2in
\bibliography{JMLRMLOSS}

\end{document}